\documentclass[aps,pra,twocolumn,longbibliography]{revtex4-1}
\usepackage{amsmath}
\usepackage{amssymb}
\usepackage{bm}
\usepackage{comment}

\newcommand\beq{\begin{equation}}
\newcommand\eeq{\end{equation}}
\newcommand\bea{\begin{eqnarray}}
\newcommand\eea{\end{eqnarray}}
\newcommand\nn{\nonumber}

\newcommand{\bra}[1]{\langle #1 |} 
\newcommand{\ket}[1]{| #1 \rangle } 

\newcommand\ketn{\ket{0}}
\newcommand\kett{\ket{t}}
\newcommand\bran{\bra{0}}
\newcommand\brat{\bra{t}}

\newcommand\Ho{\hat H}

\newcommand\dg{^\dagger}

\newcommand{\Schr}{Schr\"odinger}

\newcommand{\si}{\sigma}
\newcommand{\om}{\omega}

\newcommand\Ao{\hat A}
\newcommand\Acal{\mathcal{A}}
\newcommand\Acalo{\hat\Acal}

\newcommand\Uo{\hat U}
\newcommand\Oo{\hat O}
\newcommand\Ocal{\mathcal{O}}
\newcommand\Ocalo{\hat\Ocal}
\newcommand\sio{\hat\si}
\newcommand\siv{\bm{\si}}
\newcommand\siov{\bm{\sio}}
\newcommand\po{\hat p}
\newcommand\qo{\hat q}

\newcommand\xv{\mathbf{x}}
\newcommand\yv{\mathbf{y}}

\newcommand\Hcalo{\hat{\mathcal{H}}}

\begin{document}
\title{Unitary equivalence of \Schr\ and Heisenberg pictures survives in nonlinear quantum mechanics}
\author{Lajos Di\'osi}
\email{diosi.lajos@wigner.hu}
\homepage{www.wigner.hu/~diosi} 
\affiliation{Wigner Research Center for Physics, H-1525 Budapest 114 , P.O.Box 49, Hungary}
\affiliation{E\"otv\"os Lor\'and University, H-1117 Budapest, P\'azm\'any P\'eter stny. 1/A, Hungary}
\date{\today}

\begin{abstract}
The equivalence of \Schr\ and Heisenberg pictures is often thought to break down
in nonlinear quantum mechanics due to the absence of a state-independent unitary propagator.
We show that for Hamiltonians depending on the state through instantaneous expectation values,
the equivalence is preserved. While the unitary intertwiner becomes initial-state dependent,
it remains well-defined and ensures identical physical predictions in both pictures. 
We illustrate this with the analytically solvable examples of a nonlinear spin precession
and a mean-field harmonic force, as well as in local field theory with mean-field coupling.
\end{abstract}

\maketitle

\section{Introduction}
Nonlinear modifications of quantum mechanics have been proposed repeatedly, 
both as fundamental alternatives to the linear \Schr~dynamics and as effective 
descriptions emerging from coarse-graining, mean-field limits, feedback, or 
semiclassical couplings. Fundamental examples include the logarithmic \Schr~equation
introduced by Bia{\l}ynicki-Birula and Mycielski \cite{bialynickibirula1976}, 
and the general Hamiltonian framework of Weinberg \cite{weinberg1989testing}.
Well-known is the M{\o}ller--Rosenfeld semiclassical gravity 
\cite{moller1962,rosenfeld1963,kiefer2006quantum} whose nonrelativistic limit is
the widely discussed Newton--\Schr~equation  \cite{diosi1984,penrose1996,bahrami2014}. 
As the common feature, the  nonlinear Hamiltonians in these theories depend on the state via 
the instantaneous expectation (mean) values of certain observables.  
In Ref.~\cite{bialynickibirula1976}, there is a potential term which is proportional
to the logarithm of  the mean value of the position eigenstate projector $\ket{q}\bra{q}$.
The simplest Hamiltonian in Ref .~\cite{weinberg1989testing} assumes  Pauli spin precession 
in a field proportional to the mean value of $\sio_3$.  
In semiclassical gravity \cite{moller1962,rosenfeld1963,kiefer2006quantum}, the
Hamiltonian depends on the classical metric which depends on the mean value of the 
energy-momentum tensor of the quantized matter.  
In  the Newton--\Schr~equation \cite{diosi1984,penrose1996,bahrami2014},
the classical Newton potential is sourced by the
mean values of the spatial mass density field.

Despite this long history, a recurring view sometimes stated explicitly 
(most recently in Ref.~\cite{hsu2026relativistic}),
sometimes treated as `folklore', is that once the dynamics becomes nonlinear, 
the familiar equivalence of the \Schr~ and Heisenberg pictures becomes problematic or even meaningless.
(State dependent  Hamiltonians lead to several well-known anomalies, see, e.g., Refs .~\cite{gisin1990,diosi2016nonlinear}.
These anomalies do not occur as long as we consider only the nonlinear dyxnamics
---which is what we are doing here---without taking any quantum measurements.) 
If one insists that the Heisenberg picture must be defined via a state-independent unitary 
propagator $\Uo(t)$ that evolves the \Schr~ state $\kett$, 
then nonlinear dynamics seems to obstruct the construction at the first step: 
there is no universal $\Uo(t)$ implementing evolution for all initial states. 
Consequently, it is frequently asserted (at least implicitly) 
that in nonlinear quantum mechanics one should work only in a \Schr-like formulation
or, at best, in the interaction picture where $\Uo(t)=\exp(-it\Ho_0)$ defined by the
linear part $\Ho_0$ of the total Hamiltonian. 
The Heisenberg picture is therefore questioned as an alternative representation.
The purpose of the present short paper is to show that this conclusion is too strong.

In Sec. II, the initial-state-dependent unitary operator  $\Uo(t,\ketn)$ is introduced that evolves the 
\Schr~ state $\kett$. We prove that this unitary generates the Heisenberg operators $\Ao_t$
the same way as  $\Uo(t)$ does it in standard linear quantum mechanics. Concrete examples
of the Heisenberg picture are shown in Sec. III.

\section{\Schr~and Heisenberg dynamics, their equivalence}
Nonlinear quantum mechanics is traditionally defined in the \Schr~picture, so we
start with the nonlinear \Schr~ equation of the time-dependent statevector $\kett$:    
\beq\label{Sch}
\frac{d}{dt}\kett=-i\Ho(O_t)\kett,
\eeq
where the Hamiltonian depends on the instantaneous expectation value $O_t=\brat\Oo\kett$ of a certain
observable $\Oo$. Note that $O_t$ may stand for the expectation values of a list of different observables,
even for the continuum of expectation values of a quantum field at all locations.
For the expectation value of a generic  observable $\Ao$ (including $\Oo$), we introduce the notation $A_t$:
\beq\label{At}
A_t=\brat\Ao\kett.
\eeq
The corresponding Heisenberg dynamics, whose equivalence with \Schr's is yet to be confirmed, 
goes like this.  The quantum state is $\ketn$ constantly, and 
the observables (including $\Oo$) evolve with the nonlinear Heisenberg equation 
\beq\label{Hei}
\frac{d}{dt}\Ao_t=i[\Ho(O_t^H),\Ao_t],
\eeq
with initial condition $\Ao_0=\Ao$.
Also $\Oo_t$ satisfies the above equation, and the state dependence of the Hamiltonian is
defined by $O_t^H=\bran\Oo_t\ketn$. For the expectation value of a generic 
Heisenberg observables $\Ao_t$ (including $\Oo_t$), we introduce the notation $A_t^H$:
\beq\label{AtH}
A_t^H=\bran\Ao_t\ketn.
\eeq    
Observe that we used  notations $A_t$ in \Schr~picture and $A_t^H$ in Heisenberg picture,
respectively. The proof of $A_t=A_t^H$ is just the task along the below equivalence proof.

We are going to show that the Heisenberg picture is  unitary equivalent with the
\Schr~ picture, similarly to standard linear quantum mechanics.  
There, the picture-change unitary is state independent; 
here, for each initial condition $\ketn$ we obtain a unitary $\Uo(t;\ketn)$, 
so the equivalence holds trajectory-wise (or initial-state-wise) 
rather than by a single universal representation change.
We shall deliberately omit the $\ketn$-dependence from the notation $\Uo(t)$ since
$\ketn$ is a constant all over our calculations.   

\subsection{From \Schr~picture to Heisenberg}
To begin with the \Schr~ picture, we define the unitary operator by 
\beq\label{dUdt}
\frac{d}{dt}\Uo(t)=-i\Ho(O_t)\Uo(t),
\eeq
with $\Uo(0)=1$. The quantity  $O_t=\brat\Oo\kett$ is given via the solution of the \Schr~ eq.~\eqref{Sch} so that
$O_t$ and finally $\Uo(t)$ will depend on the initial state $\ketn$, as we previously anticipated.   
Observe now that the unitary $\Uo(t)$ `solves' the \Schr~eq.~\eqref{Sch}: 
\beq\label{Schsol}
\kett=\Uo(t)\ketn.
\eeq
Note, however, that the definition \eqref{dUdt} of $\Uo(t)$ had assumed the
knowledge of the solution $\kett$.  We define the Heisenberg observables just like we do it in the standard
linear quantum mechanics:  
\beq\label{SchH}
\Ao_t=\Uo\dg(t)\Ao\Uo(t).
\eeq
This transformation $\Ao\Rightarrow\Ao_t$ is the inverse picture  
transformation corresponding to $\ketn\Rightarrow\kett$ in Eq.~\eqref{Schsol}, like in standard quantum mechanics.
And like there, we get the same expectation values in both \Schr~ and Heisenberg pictures:
\beq\label{equiv}
A_t^H=A_t.
\eeq
This ensures the equivalence of physical predictions between the two pictures.
Last but not least, it is yet to be shown that $\Ao_t$ satisfies the Heisenberg
eq. of motion. Using eq.~\eqref{SchH} and then eq.~\eqref{dUdt}, we can write: 
\beq
\frac{d}{dt}\Ao_t=i[\Ho(O_t),\Ao_t].
\eeq
From eq.~\eqref{equiv}, we have  $O_t=O_t^H$, and this makes the above equation
coincide with the Heisenberg eq.~\eqref{Hei}.
 
\subsection{From Heisenberg picture to \Schr}
Alternatively, we can begin with the Heisenberg picture. 
We define the unitary operator by
\beq\label{dUHdt}
\frac{d}{dt}\Uo^H(t)=-i\Ho(O_t^H)\Uo^H(t)
\eeq
with $\Uo^H(0)=1$. The quantity  $O_t^H=\bran\Oo_t\ketn$ is given via the solution $\Oo_t$ of the Heisenberg eq.~\eqref{Hei} so that
$O_t^H$ and finally $\Uo^H(t)$ will depend on the initial state $\ketn$.   
This unitary `solves' the Heisenberg eq.~\eqref{Hei}: 
\beq\label{Hsol}
\Ao_t=\Uo^{H\dagger}(t)\Ao_0\Uo^H(t).
\eeq
Recall that  the definition \eqref{dUHdt} of $\Uo^H(t)$ assumes the
knowledge of the solution $\Oo_t$. 
We define the \Schr~ observables by $\Ao=\Ao_0$  and the \Schr~ state by   
\beq\label{Schsol1}
\kett=\Uo^H(t)\ketn,
\eeq
like in the standard linear quantum mechanics.  
And like there, we get the same expectation values in both \Schr~ and Heisenberg pictures,
cf. \eqref{equiv}.
Finally, we ought to inspect that the state $\kett$ in Eq.~\eqref{Schsol1} satisfies the \Schr~ equation.
Since the expectation values $O_t$ and $O_t^H$ coincide, we have $\Uo^H(t)=\Uo(t)$, 
which proves that the state $\kett$ in Eq.~\eqref{Schsol1} coincides
with the solution \eqref{Schsol} of the \Schr~ equation \eqref{Sch}.    

\section{Examples}
We demonstrate the construction of the Heisenberg picture using different nonlinear \Schr~ equations.
In two simple examples,  we provide closed analytic expressions for the Heisenberg variables and for 
the unitary map from the \Schr~picture to the Heisenberg picture. This is facilitated because the  state-dependent
$c$-number functions $O_t$ satisfy closed, analytically solvable differential equations. Analytic solutions
for the Heisenberg picture in more complex nonlinear models may also exist; here we restrict ourselves
to the simplest yet nontrivial analytically tractable examples.
Our first choice is
the nonlinear Hamiltonian $\tfrac12\langle\sio^x\rangle\sio^z$  of a two-level system (qubit), 
a variant of Gisin's example \cite{gisin1990}.
Second, a free particle is considered under the simplest state-dependent directional force $\propto\langle\qo\rangle$
as well as in the state-dependent potential $\propto(\qo-\langle\qo\rangle)^2$. This Galilean-invariant harmonic
potential was already discussed in the foundations of quantum mechanics \cite{janossy1952} 
and was rediscovered independently
in a relevant special case \cite{diosi2007,yang2013macroscopic} of the Newton--\Schr~equation. 
Our  third example is a nonrelativistic boson field with generic local mean-field coupling.

\subsection{Spin-$\tfrac12$ in  `mean-field'}
In the Pauli representation, the generic nonlinear Hamiltonian of a spin-$\tfrac12$ system is the following:
\beq
\Ho(\siv_t)=\tfrac12\bm{\om}(\siv_t)\siov,
\eeq
where the vector $\bm{\om}$ is a certain function of the Bloch vector $\siv_t=\brat\siov\kett$ at time $t$. 
The \Schr~eq.~\eqref{Sch} yields the closed rotation equation $\dot\siv=\bm{\om}(\siv)\times\siov$ 
for the Bloch vector, not solvable analytically unless we choose a specific $\bm{\om}$.
Thus, we consider the following nonlinear spin Hamiltonian \cite{gisin1990}:
\beq\label{H_spin}
\Ho(\si^x_t)=\tfrac12\om\si^x_t\,\sio^z, 
\eeq
i.e., the precession frequency about the $z$ axis is proportional to the instantaneous
$x$--component $\si^x_t=\brat\sio^x\kett$ of the Bloch vector.

In the \Schr~picture, the evolution is given by Eq.~\eqref{Sch} with $O_t\equiv \si^x_t$ and
$\Oo\equiv\sio^x$, namely
\beq\label{Sch_spin}
\frac{d}{dt}\kett=-\frac{i}{2}\om\,\si^x_t\,\sio^z\,\kett.
\eeq
Using \eqref{Sch_spin} and the commutators/anti-commutators of the Pauli matrices, 
one obtains the following closed nonlinear equations for the Bloch vector components:
\beq\label{Bloch}
\dot\si^x_t=-\om\,\si^x_t\,\si^y_t,\qquad
\dot\si^y_t=\om\,(\si^x_t)^2,\qquad
\dot\si^z_t=0.
\eeq
Hence $\si^z_t=\si^z_0$ is constant, and $(\si^x_t)^2+(\si^y_t)^2=\rho^2$ is conserved where
\beq\label{rho}
\rho\equiv\sqrt{(\si^x_0)^2+(\si^y_0)^2}=\sqrt{1-(\si^z_0)^2}\,.
\eeq
For $\rho>0$, we can parametrize the transverse components by an azimuthal angle $\phi(t)$:
\beq\label{phipar}
\si^x_t=\rho\cos\phi(t),\qquad \si^y_t=\rho\sin\phi(t).
\eeq
Substitution into \eqref{Bloch} yields a single scalar equation
\beq\label{phidot}
\dot\phi(t)=\om\rho\cos\phi(t).
\eeq
It is explicitly solvable:
\beq\label{phi_sol}
\sin\left(\phi(t)-\phi(0)\right)=\tanh(\rho\om t),
\eeq
whence $\si^x_t,\si^y_t$ follow from \eqref{phipar}.
The unitary propagator \eqref{dUdt} is in this case a $z$--rotation with a
state-dependent angle which turns out to be $-\phi(t)+\phi(0)$:
\beq\label{U_spin}
\Uo(t)=\exp\!\left[-\frac{i}{2}\bigl(\phi(t)-\phi(0)\bigr)\sio^z\right].
\eeq
Indeed, the definition \eqref{dUdt} of $\Uo(t)$, with our Hamiltonian \eqref{H_spin}, reads:
\beq\label{dUdt_spin}
\frac{d}{dt}\Uo(t)=-\frac{i}{2}\om\,\si^x_t\,\sio^z\,\Uo(t).
\eeq
Since eqs. \eqref{phipar} and \eqref{phidot}  yield $\om\si^x_t=\rho\om\cos\phi(t)=\dot\phi(t)$, 
the solution \eqref{U_spin} is confirmed, with the initial condition $\Uo(0)=1$.
The \Schr~solution is $\kett=\Uo(t)\ketn$ in agreement with \eqref{Schsol}.
Finally, the Heisenberg observables, defined by \eqref{SchH}, become the following:
\bea
\sio^x_t&=&\cos\left(\phi(t)-\phi(0)\right)\sio^x - \sin\left(\phi(t)-\phi(0)\right)\sio^y\nn\\
\sio^y_t&=&\cos\left(\phi(t)-\phi(0)\right)\sio^y +\sin\left(\phi(t)-\phi(0)\right)\sio^x\nn\\
\sio^z_t&=&\sio^z
\eea
The Heisenberg observables $\siov_t$ satisfy the equivalence condition \eqref{equiv}: 
\beq
\bran\,\siov_t\,\ketn=\brat\,\siov\,\kett=\siv_t~,
\eeq
and the standard commutator-anticommutator relationships of the Pauli matrices hold
for the components of $\siov_t$ in the Heisenberg picture as well.

\subsection{`Mean-field' directional force}
Consider a free particle ($m=1$) of momentum $\po$ and position $\qo$, 
subject to a directional force proportional to its mean position $q_t = \brat\qo\kett$.
The state-dependent Hamiltonian reads:
\beq\label{H_force}
\Ho(q_t)=\tfrac12\po^2+\om^2 q_t \qo. 
\eeq
This time we start in the Heisenberg picture.
The nonlinear Heisenberg equation \eqref{Hei} yields the equation of motion of the canonical variables:   
\beq\label{Hei_force}
\dot{\qo}_t = \po_t,~~~~ \dot{\po}_t = -\om^2 q_t. 
\eeq
The expectation values obey the closed canonical equations of a classical harmonic oscillator 
of frequency $\om$. The elementary solutions are
\beq\label{qtpt}
q_t = q_0 \cos \om t + \frac{p_0}{\om} \sin \om t, \qquad p_t = p_0 \cos \om t - \om q_0 \sin \om t,
\eeq
depending on the initial values $q_0 = \bran\qo\ketn$ and $p_0 = \bran\po\ketn$.
Now we can solve the Heisenberg equations \eqref{Hei_force}.
They are driven by the $c$-number force $-\om^2 q_t$ whose time-dependence
is known from \eqref{qtpt}. Integrating them yields the explicit 
Heisenberg operators:
\beq\label{qotpot}
\po_t = \po +(p_t-p_0) , \qquad \qo_t = \qo + t \po + (q_t-q_0-p_0t).
\eeq
One can directly verify 
the canonical commutator $[\qo_t,\po_t]=i$ in the Heisenberg picture.
To derive the unitary $\Uo(t)$, we can skip the defining equation \eqref{dUdt} because
$\Uo(t)$ is already defined by the above unitary transformation of the canonical variables
$(\qo,\po)$  from the \Schr~picture to $(\qo_t,\po_t)$ in the Heisenberg picture. 
The propagator $\exp(-it\Ho)=\exp(-\tfrac12 it\po^2)$ acts first,
then the position and momentum operators are shifted by  $q_t-q_0-p_0t$ and $p_t-p_0$ respectively.
The unitary $\Uo(t)$ ---that generates the Heisenberg observables via eq.~\eqref{SchH}
as well as the \Schr~state via eq.  \eqref{Schsol}--- is given by:
\beq\label{U_force}
\Uo(t) = \exp\bigl[-i (p_t-p_0) \qo\bigr] \exp\bigl[i(q_t-q_0-p_0t)\po\bigr] \exp(-it\Ho),
\eeq
up to a $c$-number phase. 
This $\Uo(t)$ depends on the initial state $\ketn$ only through the initial expectation values 
$q_0, p_0$ of the solutions \eqref{qotpot}. 
The construction is valid for any initial state $\ketn$, demonstrating that 
unitary equivalence is preserved despite the non-trivial nonlinearity.

It is even easier to construct the Heisenberg picture for a slightly different nonlinear \Schr~ equation:
\beq\label{H_NSE}
\Ho(q_t)=\frac12\po^2+\frac12\om^2 (\qo-q_t)^2, 
\eeq
where the directional force is centered around the instantaneous mean position. 
We exploit the Galilean invariance of the nonlinear dynamics.
It coincides with the standard (linear) dynamics of the central harmonic oscillator in
the specific inertial frame where $q_0=p_0=0$, the nonlinear Hamiltonian becomes
the linear Hamiltonian $\Ho_0=\tfrac12(\po^2+\om^2\qo^2)$ of the central
harmonic oscillator. The Heisenberg picture is generated by the unitary $\exp(-it\Ho_0)$.
We can transform it into the general frame by Galilean transformations of the coordinate
and momentum, yielding
\bea
\qo_t&=&\qo\cos\om t+\frac{\po}{\om}\sin\om t+q_0+p_0t\nn\\
\po_t&=&-\om\qo\sin\om t+\po\cos\om t+p_0\\
\Uo(t)&=&\exp\bigl[-i p_0 \qo\bigr] \exp\bigl[i(q_0+p_0t)\po\bigr] \exp(-it\Ho_0).\nn
\eea

\subsection{Local mean-field coupling}
In the  \Schr~ picture, let $\Acal(\xv)$ and $\hat{\mathcal{B}}(\xv)$ be local observables of a boson field theory,
satisfying the locality (microcausality) condition: 
\beq\label{loc}
[\Acalo(\xv),\hat{\mathcal{B}}(\yv)]=0,~~~~ (\xv\neq\yv).
\eeq 
Consider the expectation value of a certain local field:
\beq
\Ocal_t(\xv)=\brat\Ocalo(\xv)\kett
\eeq
which enters the local field $\Hcalo(\xv,\Ocal_t(\xv))$ of Hamiltonian density, yielding the
nonlinear Hamiltonian:
\beq\label{H_lmf}
\Ho[\Ocal_t]=\int\Hcalo(\xv,\Ocal_t(\xv))d\xv.
\eeq
The nonlinear \Schr~ equation reads
\beq
\frac{d}{dt}\kett=-i\Ho[\Ocal_t]\kett .
\eeq
For a given $\ketn$, there is a unique solution $\kett$, and hence a unique solution $\Ocal_t$, and
a unique solution $\Ho[\Ocal_t]$. Therefore,  there exists a unique solution of \eqref{dUdt} whose
unitarity is most clearly displayed by the time-ordered explicit form: 
\beq
\Uo(t)=T\exp\left(-i\int_0^t\Ho[\Ocal_\tau]d\tau\right).
\eeq 
Remember, the correct notation would be $\Uo(t;\ketn)$, just the initial-state-dependence 
is suppressed in our notation.
Using this initial-state-dependent unitary, we define the Heisenberg fields:
\beq
\Acalo_t(\xv)=\Uo\dg(t)\Acalo(\xv)\Uo(t).
\eeq

No matter that the unitary map depends on the initial state $\ketn$,
the map preserves the algebra of observables: 
\beq
[\Acalo_t(\xv),\hat{\mathcal{B}}_t(\yv)]=[\Acalo(\xv),\hat{\mathcal{B}}(\yv)].
 \eeq
It follows that  the Heisenberg fields preserve 
the locality (microcausality) \eqref{loc} of the \Schr~ fields.

\section{Summary, and closing remarks}
We have shown that the standard equivalence between the \Schr~ and the Heisenberg pictures 
survives in nonlinear quantum mechanics, provided the nonlinearity enters through expectation values.
The key is to recognize that while a universal, state-independent unitary intertwiner $\Uo(t)$ 
between the two pictures no longer exists,
a trajectory-wise unitary $\Uo(t;\ketn)$ can always be constructed for any given initial state. 
(The equivalence of the interaction picture follows by similar arguments.)
The time-derivative $d\Uo\!/\!dt$ satisfies the same differential equation as in the linear theory.  
As demonstrated by our examples, the nonlinear Heisenberg equations remain
an equivalent and transparent alternative to the nonlinear nonlinear \Schr~ equations. 

This work was motivated originally by the author's disagreement with the 
statement in Ref.~\cite{hsu2026relativistic} that nonlinear
dynamics does not preserve  the microcausality (locality) condition of 
quantum fields because the evolution is non-unitary. The contrary
argument \cite{diosi2026comment}, which states  that  the evolution of fields is initial-state-dependent \emph{unitary},
also ensures the equivalence of the \Schr\  and Heisenberg pictures.
Very recently, Ref.~\cite{liu2025semiclassical} has already recognized this equivalence in
a particular nonlinear quantum model, but the authors did not need to set out to prove that
as a general rule.  

\acknowledgments
This research was supported by the EU COST Actions (Grants
CA23115, CA23130). The author thanks an AI assistant for helpful discussions.

\bibliography{diosi2025}{}
\end{document}